\documentclass[showpacs,showkeys,floatfix]{revtex4}
\usepackage[dvips]{graphicx}
\begin{document}

\title{Soft modes and thermal transport in guest-host crystalline structures.}

\author{A.R.Muratov$^{1}$}
\email{muratov@ogri.ru}
\affiliation{$^1$ Institute for Oil and Gas Research, Moscow, Russia, 
and Institute Laue-Langevin, F-38042, Grenoble, France;}
\author{E.I.Kats$^2$}
\email{kats@ill.fr}
\affiliation{$^2$ Institute Laue-Langevin, F-38042, Grenoble, France and L.D. Landau Institute for Theoretical Physics, Moscow, Russia.}
\author{H.Schober$^3$}
\email{schober@ill.fr }
\affiliation{$^3$ Institute Laue-Langevin, F-38042, Grenoble, France;}

\date{today}

\begin{abstract}
We propose a simple phenomenological model describing vibrational dynamics in a class
of materials (like clathrates), which are crystalline compounds with closed cavities in their structures.
In the spirit of a minimalist approach, our model includes only the minimum of ingredients necessary to capture the
generic features of this class of compounds as brought to light by experiments. 
We consider only two kinds of particles: strongly coupled host atoms, which form the structural cavities, and guest atoms confined in the cavities.
In spite of the confinement, for relatively small amplitude vibrations of the guest atoms, 
their interaction with the host particles is assumed to be weaker than between the host atoms. 
We calculate self-consistently the vibrational mode (phonon) 
line broadenings for the empty
and filled (by the guest atoms) cavities.
We show that the soft phonon dispersion sheet, which appears for weakly bonded guest atoms,
yields to large mode broadening via three waves anharmonic mode coupling. 
In turn the Umklapp part of this broadening reduces drastically the phonon thermoconductivity coefficient.
We conclude that for a system with a soft mode component in the 
vibrational spectrum phonon broadening is always larger  and thermal conductivity is always appreciably smaller 
than for a system without this soft mode (provided all other characteristics are the same). 
Although our model is minimal, in the sense of ignoring the complexity of actual guest-host structures, 
when properly interpreted, it can yield quite reasonable values for a variety of measurable quantities.

\end{abstract}


\maketitle

\section{Introduction}
Among the various types of so-called guest - host systems (see e.g., the recent review \cite{KS} containing
numerous relevant references) one of the simplest type can be constructed from two kinds of particles,
the strongly coupled host atoms forming structural cavities, and guest atoms, which can be captured in the cavities.
The term guest atoms already indicates that they should not play a strong structural role although actual host structures 
may become unstable in the absence of the guest.
We may, therefore consider the situation of guest particles that are confined within the cavities, but apart from this, 
are only weakly interacting with the host atoms.
The interest to study these systems reflects both, their potential practical importance (related to the possibility
of selectively tuning mechanical, electrical, or thermal properties), and the associated theoretical challenges.
However, complexity of composition of existing guest-host systems (like clathrate hydrates, or skutterudites)
that underlies the rich physical properties, poses a formidable problem to theoretical and computational modeling efforts.
Although a number of detailed experimental investigations and sophisticated calculations and numerical computations
have been published over the last few years (see, e.g., papers \cite{SM97} - \cite{KJ08}, which partially will be commented
upon in our paper, and much more can be easily added
to this list), there is still a clear need for a simple (but yet non-trivial) theoretical model with predictions,
which can be confronted with experimental observations. The fact (perhaps a bit taste-dependent) is that more or less realistic
and complete theoretical or numerical investigations of existing guest - host systems (with about 50 atoms per elementary
cell) do not yield simple intuitive (yet based on firm physical principles) explanations of the vibrational excitation broadening and the 
considerable reduction of thermal conductivity upon filling of the host structural cavities with guest atoms.
In spite of the fact that the basic concepts describing vibrational excitations and thermal transport were born long ago and can be found in the textbooks
\cite{ZI60}, \cite{KI68}, many questions of principle  remain to be settled for guest-host systems  (see e.g.\, the discussions
about Einstein rattlers in the very recent papers \cite{KJ08}, \cite{CA08}, or about electric dipole force contributions
in \cite{NK08}). In part this frustrating situation is just due to the lack of a simple and tractable analytical model.
Our motivation for adding one more paper to the topic is precisely to propose such a model.

The outline of this article is as follows. In Section \ref{bas} we formulate our model of a simplest guest - host system
and calculate dispersion curves. In Section \ref{mod} we derive self-consistent equations for vibrational mode broadening, which 
are solved 
numerically in Section \ref{num}. Based on this solution we compute also thermoconductivity coefficients.
Section \ref{con} is devoted to a discussion, and summary of our main results.

\section{Preliminaries. }
\label{bas}
We shall recall certain basic facts about the guest-host systems.
A generic structural feature of this class of solids is that their lattices are formed by the strongly interacting host atoms or molecules. 
On the contrary the guest molecules can be located only in the cages formed within the host structure.
The basic interaction between the guest molecule and the host lattice is generally weaker than between the host molecules,
i.e.\, the elastic force controlling the guest particle's vibrations is somehow small.
However, for larger displacements the interaction between the guest and host lattice becomes stronger holding the guest 
particles firmly within the cage. 
The guest-host interaction can in this sense be assimilated to an inverted hard-core potential, soft in the center and hard outside.
The weak elastic coupling property leads to the presence of low-frequency or soft dispersion sheets in the system, which correspond
to the vibrations of the guest molecules within the cage.
For the sake of simplicity, let us sketch first a one-dimensional cartoon (see Fig. 1) corresponding to our model structure (later on in the next 
section we will do a proper 3D calculation). 
Essential for our model is that the guest-host structure elementary cell contains two kinds of particles: structural atoms $H$ (host) and guest atoms $G$.
In equilibrium the $H$ atoms occupy positions $x_H=2n+1$, and there are only nearest neighbor interactions between the $H$ atoms.
Guest atoms $G$ are located between the atoms $H$ at the equilibrium positions $x_G=2n$ and they
interact only with the host atoms forming their cages. 

\begin{figure}
\centering
\includegraphics[width=80mm]{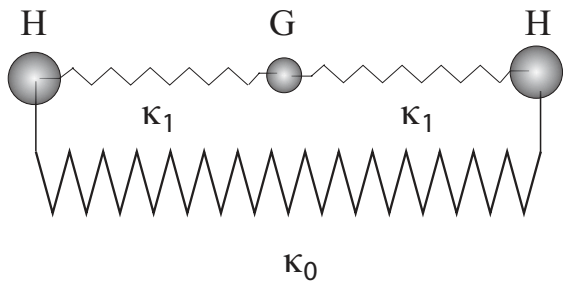}
\caption{Cartoon 1D guest-host structure}
\label{f1}
\end{figure}

The elastic energy for such a system can be written as
\begin{equation}
E=\frac 12\sum_n \left( \kappa_0 (u_{2n+1}-u_{2n-1})^2\ +\kappa_1 (u_{2n}-u_{2n-1})^2+\kappa_1(u_{2n}-u_{2n+1})^2 \right) \ , 
\label{1}
\end{equation}
where $u_n$ is a displacement of atom $n$, $\kappa_0$ and $\kappa_1$ are the host -host and guest -host force constants.
Soft guest modes require that $\kappa_1 < \kappa_0$. 

For the ease of algebra we are using dimensionless units where both masses ($H$ and $G$ atoms) are set equal to $2$.
The dynamical equations describing the eigenmodes and eigenvectors of this system read as
\begin{eqnarray}
&&2\frac{d^2 u_{2n+1}}{dt^2}=\kappa_0(2u_{2n+1}-u_{2n-1}-u_{2n+3})
+\kappa_1(2u_{2n+1}-u_{2n}-u_{2n+2}) \ ,
\nonumber \\
&&2\frac{d^2 u_{2n}}{dt^2}=\kappa_1(2u_{2n}-u_{2n-1}-u_{2n+1})\ .
\label{2}
\end{eqnarray}

\begin{figure}
\centering
\includegraphics[width=80mm]{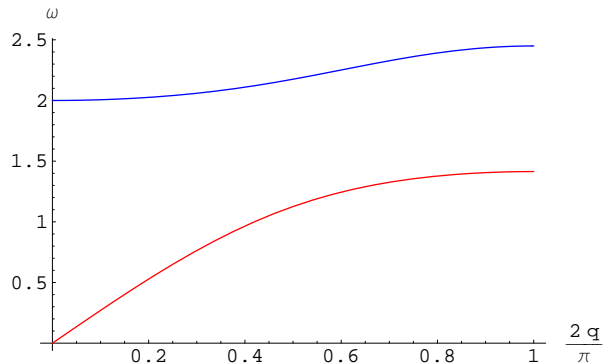}
\caption{Spectrum of eigenmodes for the elastic constants $\kappa_0=\kappa_1=2$ (1D model).}
\label{f2}
\end{figure}

\begin{figure}
\centering
\includegraphics[width=80mm]{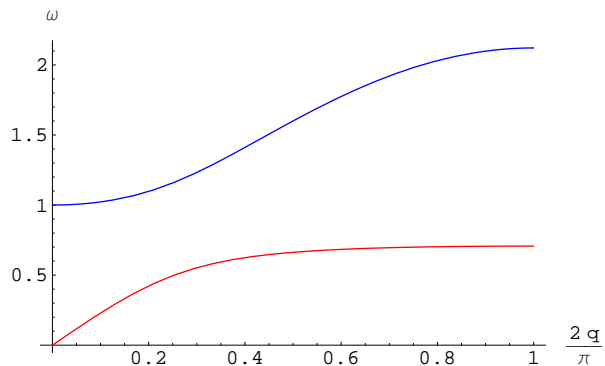}
\caption{Spectrum of eigenmodes for the elastic constants $\kappa_0=2,\kappa_1=0.5$ (1D model).}
\label{f3}
\end{figure}

Using the standard Ansatz for the excitations of the form 
\begin{eqnarray}&& u_{2n+1}(t)=a_1\,\exp(-i\omega t +iq(2n+1)),\nonumber \\
                              && u_{2n}(t)=a_2\,\exp(-i\omega t +i2qn) \nonumber
\end{eqnarray}
                               this system of equations reduces to
\begin{eqnarray}
&&(2\kappa_0\sin^2(q)-\omega^2+\kappa_1)a_1 -\kappa_1\cos(q)a_2 =0 \ ,
\nonumber \\
&&-\kappa_1\cos(q)a_1 +(\kappa_0-\omega^2)a_2=0 \ .
\label{3}
\end{eqnarray}
Nontrivial solutions exists at $\omega=w_1(q)$ (acoustic branch) or $\omega=w_2(q)$ (optical branch) given by the relation
\begin{equation}
w_{1,2}^2(q)=\kappa_1+\kappa_0\sin^2(q)\mp \sqrt{\kappa_1^2\cos^2(q)+\kappa_0^2\sin^4(q)}
\ .
\label{4}
\end{equation}
The spectrum of eigenmodes as defined by equation (\ref{4}) is presented in Figures \ref{f2} and \ref{f3}. 
In Fig. \ref{f2} the force constants are choose as $\kappa_0=\kappa_1=2$.
In Fig. \ref{f3} $\kappa_0=2$ as before but the guest-host force constant is appreciably smaller, $\kappa_1=0.5$. 
Calculating the eigenvectors of the system (\ref{3}) 
it is easy to show, that the region of small wave-vectors within the acoustic branch corresponds to in-phase oscillations of both host and guest atoms.
For the case presented in Fig. \ref{f3}, the region of large wave-vectors of the 
lower "soft" branch corresponds to oscillations of dominantly the guest atoms.
The flat part of this branch resembles that of Einstein-like independent rattling oscillations, which are widely discussed in the literature 
\cite{KS} - \cite{CA08}. 
Despite a certain similarity we would like to stress that, in our model the flatness of the branch reflects the soft dynamics of the guest atoms, and 
not decoupled dynamics of Einstein oscillators \cite{HG05}.

As can be easily demonstrated the property controlling the form of the spectrum (\ref{4}) is the minimal value 
$\Delta ^2_{min}$ of the gap $\Delta ^2(q) = w_2^2(q) - w_1^2(q)$ between the two branches.
It reads as
\begin{equation}
\Delta _{min}^2 = 2\kappa _1\left (1 - \frac{\kappa _1^2}{4\kappa _0^2}\right )^{1/2} 
\ 
\label{h1}
\end{equation}
(cp. with the direct absolute gap defined as $(w_1^{max})^2 - (w_2^{min})^2 = \kappa _1$).
Therefore, as had to be expected the ratio $\kappa _1/\kappa _0$ is  (up to a scaling factor) the only relevant parameter in our model.
Upon decreasing the ratio $\kappa _1/\kappa _0$, one gets a more extended ''flat'' part of the spectrum (what is
supposed to be a ''rattler''-like decay channel increasing mode broadening). Even more, the reader may be surprised
to learn that for sufficiently small $\kappa _1/\kappa _0 \leq 1/7$ the large $q$ (short wavelength) upper branch group velocity, could become
larger than the long wavelength acoustic mode (sound) speed. 
In these unusual conditions one would have phonon broadening of this branch via interactions to the lowest lying long wavelength acoustic mode, which
for normal downward sloping modes are forbidden due to momentum and energy conservation rules. 
However, upon decreasing the $\kappa _1/\kappa _0$ ratio, we simultaneously increase
the gap (\ref{h1})  and therefore the mode anharmonic coupling (see the next section \ref{mod}
for more details) to provide comparable with a larger value of the $\kappa _1/\kappa _0$ broadening, should be much stronger.

Thus for strong broadening there is a certain optimum value of the ratio $\kappa _1/\kappa _0$ , not too small (to have a reasonably
strong mode coupling) and not too large (to have a reasonably extended flat spectrum). 
Instead of a full systematic study of these issues we perform in this work a few numerical estimates to narrow down the focus of our problem. 
In what follows just this optimal range of the model parameters will be investigated. 

Summarizing up to this point, for our minimal model in the harmonic approximation we have at our disposition one physical 
parameter $\kappa _1/\kappa _0$ that allows us to tune two phenomena (namely, mode flatness and mode softness, however not independently) 
that contribute  to vibrational broadening and thermoconductivity reduction. 
One might generalize the model to include also guest - guest coupling. However, because
it is not a major goal of our work to achieve agreement between the results obtained
with our phenomenological model and experimental measurements, we defer the generalization to future publications.
Nevertheless since the present understanding of mechanisms leading to thermoconductivity reduction in guest - host systems is incomplete,
even our minimal model may be an appropriate tool for working out typical trends that in principle can be tested in experiments.

\section{Vibration excitation spectrum}
\label{mod}

Phonons are crucial for the understanding not only of vibrational properties but equally of a material's electric and thermal transport.
They are also at the heart of a broad range of interesting electro-mechanical effects. 
In particular for the transport properties phonon decay plays an important role.
The problem of phonon decay has a rather long history, and the phenomenon has been treated in quite some depth in order to rationalize 
neutron scattering data concerning collective excitations. 
Theoretical concepts on anharmonic vibrational modes in crystalline materials have been developed in the seminal papers
by Peierls \cite{PI29}, Pomeranchuk \cite{PO43}, Herring \cite{HE54}, and Klemens \cite{KL58} (see also 
early works on the decay of an optical phonon into two acoustic phonons via anharmonicities \cite{OR66}, \cite{KL66},
and a more recent publication \cite{HK93}). 
The short version of the formal machinery starts as its initial step to uncouple the coupled
lattice oscillators, and then to express the Hamiltonian for the vibrating lattice as a sum of separate oscillators.
Free phonons (in our model with the dispersion laws (\ref{4})) result from the quantization of these oscillators.
In the harmonic (quadratic over lattice displacements) approximation there is no interaction between phonon states, and
an arbitrary number of phonons may be present in each mode. 
The anharmonic terms present the processes involving
three and more phonons describing their decay or coalescence.
To find the broadening for the phonon dispersions calculated  in the previous section \ref{bas} in the harmonic approximation
one has to include phonon-phonon interaction self-consistently. The diagram technique is an appropriate tool to treat such a problem.
We skip almost all general technical details, and focus here on ingredients indispensable for
our calculations. 

To be more specific we take into account only three wave interactions, because usually these interactions
are responsible for the mode broadening \cite{PO43} - \cite{KL58}. 
In the symmetric case these three phonon couplings are vanishing, we thus have to introduce asymmetry into our simple model in order to 
achieve non-zero three-wave interaction. The asymmetry can be related to a off-center position of the guest atom within the cage,
or merely non symmetric spring constants between the guest atom and the cavity walls.
In this case the anharmonic higher order couplings do not introduce new physics, but only renormalize parameter values of the coupling terms.
Since these spring constants are not well known, we regard their values 
as phenomenological coefficients. In this sense our model should be treated as a working hypothesis. 
Comparison of the qualitative predictions resulting from this hypothesis with experimental or 
microscopical numerical observations will show whether and when this hypothesis is justified.
Note also that it is a priori no problem to include also fourth order anharmonic couplings into 
the same self-consistent treatment. This will not affect our qualitative and semi-quantitative conclusions much (albeit at the expense of rapidly
increasing complexity). We do believe that with the aim of answering first simple questions transparency is worth a few simplifications.
Guided by this principle we consider in what follows only three wave couplings and only high temperatures $T > T_D$, where
$T_D$ is the Debye temperature.

In canonical phonon variables the Hamiltonian of the system can be presented as
\begin{equation}
H=\int \frac {dq}{2\pi }\, \left( w(q)a(q)a^*(q) +\int \frac {dp}{2\pi }\, \frac V6(a(q)a(p)a^*(q+p)+c.c.)\right)
\ .
\label{5}
\end{equation}
The vertex $V$ for the interaction of three phonons with quasi wave-vectors ${\bf q}$, ${\bf p}$ and ${\bf q}+{\bf p}$ 
and frequencies $\omega$, $\nu$ and $\omega +\nu$ is proportional to the product
\begin{equation}
V=\frac {qp(q+p)}{\sqrt{\omega\nu (\omega+\nu)}}\tilde V
\ .
\label{6}
\end{equation}
For simplicity we suppose that the parameter $\tilde V$ is a constant.
For the sake of the skeptical reader we should note, that if the above mentioned asymmetry of the guest - host
interactions or/and equilibrium positions is somewhat small, the three-wave interaction vertex $V$ has to be small as well.
It is not to say that guest-host vibrational mode coupling is small. 
Indeed, as a guest atom approaches the wall of the cavity, repulsive interactions become
dominant and the host atoms in the vicinity of the wall are pushed back to decrease the unfavorable interactions. Consequently,
host-guest vibrations are strongly coupled, even though the direct attractive potential between the guest and the host lattice is small.
To avoid too many unknown parameters we are not introducing explicitely this asymmetry coefficient in the (\ref{5}), (\ref{6}).
All the more that in this study we are predominantly concerned with only relative
values of phonon mode broadenings and thermoconductivities for pure host and guest-host systems.

\begin{figure}
\centering
\includegraphics[width=80mm]{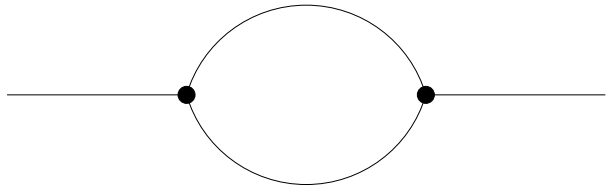}
\caption{One-loop diagrams for phonon mode broadenings}
\label{diag}
\end{figure}

The peculiarities of the phonon mode broadenings are related to the existence of the soft mode, and these effects
are determined by inter-band coupling (i.e., the decay of one optical or acoustic phonon into two
acoustic phonons, or into one acoustic and one optic phonon, and inverse coalescence processes, satisfying the simple
selection rule that the created phonon must lie in a higher branch than at least one of the destroyed phonons)
integrated over the full range of 
wave vectors.
In the one-loop approximation, 
the broadening of the phonon spectra can be determined from 
diagrams of the type schematically shown in Fig. \ref{diag},
where the lines describe phonons which decay or coalescence.
We take into account all possible processes involving three phonons (not only inter-band ones, mentioned above).
Calculating these diagrams we end with
the 
following system of equations
to find 
the broadening of the phonon spectra  
\begin{eqnarray}
&&\gamma_1(q)=\int_{-\pi/2}^{\pi /2} \frac {dp}{2\pi }\, V^2 \Bigl( 
\frac{(\gamma_1(p)+\gamma_1(q+p))(n(w_1(p))-n(w_1(q+p)))}
{(w_1(q)+w_1(p)-w_1(q+p))^2+(\gamma_1(p)+\gamma_1(q+p))^2}
\nonumber \\
&&+\frac{(\gamma_1(p)+\gamma_1(q-p))(1+n(w_1(p))+n(w_1(q-p)))/2}
{(w_1(q)-w_1(p)-w_1(q-p))^2+(\gamma_1(p)+\gamma_1(q-p))^2}
\nonumber \\
&&+\frac{(\gamma_1(p)+\gamma_2(q+p))(n(w_1(p))-n(w_2(q+p)))}
{(w_1(q)+w_1(p)-w_2(q+p))^2+(\gamma_1(p)+\gamma_2(q+p))^2}\Bigr)
\ , \nonumber \\
&&\gamma_2(q)=\int_{-\pi /2}^{\pi /2} \frac {dp}{2\pi }\, V^2 \Bigl(
\frac{(\gamma_1(p)+\gamma_2(q+p))(n(w_1(p))-n(w_2(q+p) ))}
{(w_2(q)+w_1(p)-w_2(q+p))^2+(\gamma_1(p)+\gamma_2(q+p))^2}
\nonumber \\
&&+\frac{(\gamma_1(p)+\gamma_1(q-p))(1+n(w_1(p))+n(w_1(q-p)))/2}
{(w_2(q)-w_1(p)-w_1(q-p))^2+(\gamma_1(p)+\gamma_1(q-p))^2}
\nonumber \\
&&+\frac{(\gamma_1(p)+\gamma_2(q-p))(1+n(w_1(p))+n(w_2(q-p) ))}
{(w_2(q)-w_1(p)-w_2(q-p))^2+(\gamma_1(p)+\gamma_2(q-p))^2}\Bigr)
\ .
\label{7}
\end{eqnarray}
Here subscripts 1 and 2 stands for acoustic and optical branches of the spectrum, $n(\omega)$ 
is the Bose factor, $n(\omega)=1/(\exp(h\omega/k_B/T)-1)$.
To stay in contact with experiments we consider only the high temperature case 
when the Bose factor turns into the Boltzmann law $n(\omega)= k_BT/(h\omega) \gg 1$.
One more comment is in order here. Of course physical phonon damping is determined by the functions $\gamma _1$ and
$\gamma _2$ calculated at the actual mode dispersion laws (\ref{4}) (so-called ''on-shell'' phonon broadening).
However as an intermediate step of calculations, when we wish to include self-consistently
the effects of phonon damping, ''off-shell'' functions $\gamma $ can enter the above integral equations (\ref{7}).
We check that our self-consistent equations are stable with respect to off-shell virtual phonons, and thus
their contributions are irrelevant for our work.

Then, the equations (\ref{7}) can be rewritten as
\begin{eqnarray}
&&\gamma_1(q)=\frac{\lambda q^2}{w_1(q)}\int_{-\pi /2}^{\pi /2} dp\, f_1^2(p)\Bigl( 
\frac{f_1^2(s)(\gamma_1(p)+\gamma_1(s))(w_1(s)-w_1(p))}
{(w_1(q)+w_1(p)-w_1(s))^2+(\gamma_1(p)+\gamma_1(s))^2}
\nonumber \\
&&+\frac{f_1^2(s)(\gamma_1(p)+\gamma_1(s))(w_1(s)+w_1(p))/2}
{(w_1(q)-w_1(p)-w_1(s))^2+(\gamma_1(p)+\gamma_1(s))^2}
\nonumber \\
&&+\frac{f_2^2(s)(\gamma_1(p)+\gamma_2(s))(w_2(s)-w_1(p))}
{(w_1(q)+w_1(p)-w_2(s))^2+(\gamma_1(p)+\gamma_2(s))^2}\Bigr)
\ , \nonumber \\
&&\gamma_2(q)=\frac{\lambda q^2}{w_2(q)}\int_{-\pi /2}^{\pi /2} dp\, f_1^2(p)\Bigl(
\frac{f_2^2(s)(\gamma_1(p)+\gamma_2(s))(w_2(s)-w_1(p))}
{(w_2(q)+w_1(p)-w_2(s))^2+(\gamma_1(p)+\gamma_2(s))^2}
\nonumber \\
&&+\frac{f_1^2(s)(\gamma_1(p)+\gamma_1(s))(w_1(s)+w_1(p))/2}
{(w_2(q)-w_1(p)-w_1(s))^2+(\gamma_1(p)+\gamma_1(s))^2}
\nonumber \\
&&+\frac{f_2^2(s)(\gamma_1(p)+\gamma_2(s))(w_2(s)+w_1(p))}
{(w_2(q)-w_1(p)-w_2(s))^2+(\gamma_1(p)+\gamma_2(s))^2}\Bigr)
\ ,
\label{8}
\end{eqnarray}
where $\lambda= T{\tilde V}^2/(2\pi )$, $f_{1,2}(p)=|p|/w_{1,2}(p)$ and $s=q+p$. 
To take into account Umklapp processes it is necessary to extend the functions 
$w(q)$ and $\gamma(q)$ into the region $[-3\pi /2\, , \,3\pi /2]$ as a periodical repeat of the first Brillouin zone $[-\pi/2 \, , \, \pi /2]$.
Equations (\ref{8}) give the imaginary part of the one-loop correction to the bare Green 
functions, which determines the decrements. For simplicity we neglect the real part of this correction, 
which only changes the bare dispersion laws \ref{4}, and treat the system (\ref{8}) self-consistently in the 
functions $\gamma_1(q)$ and $\gamma_2(q)$.
The system can be solved numerically, and as a result we can calculate the broadenings of the two
branches (\ref{4}) as a function of parameters $\lambda$ and the ratio $\kappa_1/\kappa_0$. 

These self-consistent equations (\ref{8}) are derived yet for the one-dimensional model shown in Fig. 1,
and the result overestimates the contributions from small wave-vectors (although it is worth noting that as
already pointed out by Peierls \cite{PI29} for phonon-phonon interactions the major physical difference is between one dimension
and higher. Three-phonon interactions in one dimension are severely restricted and usually very weak). 
However, the derivation can be easily generalized to three dimensional systems.
Because it is not a major goal of this work to achieve quantitative agreement between theoretical results obtained
with this phenomenological model and experimental data, we restrict the three-dimensional consideration to 
the same vibrational modes (one acoustic and one optic dispersion sheet). 
Note also that for finite temperature ($T > T_D$) three wave broadening, the integrals entering the self-consistent equations
 for the broadenings, only weakly depend on the region of wave vectors in 
the vicinity of the boundary of the Brillouin zone. Therefore,
aiming only at a qualitative description of experimental data, in the first approximation one can perform this integration over an 
appropriately (i.e., with equal volumes) chosen 
spherical volume. Combining everything together we end up with the following three dimensional self-consistent equations
for the mode broadening in the guest-host system. 
\begin{eqnarray}
&&\gamma_1(q)=\frac{\lambda q^2}{w_1(q)}\int_{-1}^1 dt \int_0^{\pi/2}dp\, p^2f_1^2(p)
\nonumber \\
&&\Bigl( \frac{f_1^2(s)(\gamma_1(p)+\gamma_1(s))(w_1(s)-w_1(p))}
{(w_1(q)+w_1(p)-w_1(s))^2+(\gamma_1(p)+\gamma_1(s))^2}
\nonumber \\
&&+\frac{f_1^2(s)(\gamma_1(p)+\gamma_1(s))(w_1(s)+w_1(p))/2}
{(w_1(q)-w_1(p)-w_1(s))^2+(\gamma_1(p)+\gamma_1(s))^2}
\nonumber \\
&&+\frac{f_2^2(s)(\gamma_1(p)+\gamma_2(s))(w_2(s)-w_1(p))}
{(w_1(q)+w_1(p)-w_2(s))^2+(\gamma_1(p)+\gamma_2(s))^2}\Bigr)
\ , \nonumber \\
&&\gamma_2(q)=\frac{\lambda q^2}{w_2(q)}\int_{-1}^1 dt \int_0^{\pi/2}dp\, p^2f_1^2(p)
\nonumber \\
&&\Bigl( \frac{f_2^2(s)(\gamma_1(p)+\gamma_2(s))(w_2(s)-w_1(p))}
{(w_2(q)+w_1(p)-w_2(s))^2+(\gamma_1(p)+\gamma_2(s))^2}
\nonumber \\
&&+\frac{f_1^2(s)(\gamma_1(p)+\gamma_1(s))(w_1(s)+w_1(p))/2}
{(w_2(q)-w_1(p)-w_1(s))^2+(\gamma_1(p)+\gamma_1(s))^2}
\nonumber \\
&&+\frac{f_2^2(s)(\gamma_1(p)+\gamma_2(s))(w_2(s)+w_1(p))}
{(w_2(q)-w_1(p)-w_2(s))^2+(\gamma_1(p)+\gamma_2(s))^2}\Bigr)
\ ,
\label{9}
\end{eqnarray}
where $s=\sqrt{q^2+p^2-2qpt}$.

The self-consistency equations (\ref{9}) include as pointed out above only two modes: one optic and one acoustic (the latter one
is longitudinal by its construction). 
Thus, we have in particular neglected two transverse acoustic modes. Lax et al., \cite{LH81} have shown that a given acoustic
phonon cannot decay into other modes with higher velocity at any order in the anharmonicity. 
For the lowest-lying acoustic phonon mode one then expects anomalously long lifetimes. 
Similarly the phonon Umklapp processes have negligible effect on the heat flux
carried by the low-energy phonons with small wave vectors. The latter would lead to 
very high thermal conductivity. For this reason, calculating the phonon broadening and thermal conductivity,
we cut-off corresponding integrals (see below) at a certain small wave vector. Being interested in the only relative
effects upon inserting guest atoms into the host lattice,
this, and some other
approximations made above driven by pure desire to make formulae simpler and to clarify physical ideas, do not affect
our conclusions on the thermoconductivity reduction qualitatively. 
Moreover all these assumptions are easy to relax, if needed, for example other phonon-scattering mechanisms,
providing finite broadening for long-wave phonons, can be considered on the same footing.

\section{Numerical results}
\label{num}

The self-consistent integral equations (\ref{9}) are our main results and they are ready for further inspection
to find their solutions. Solving the system (\ref{9}) we find the broadening of the spectrum. 
Luckily for us the solutions to these  rather complicated looking equations (e.g., they are not symmetric with respect
to $ 1 \to 2$ exchange, as one could expect since there is no symmetry relation between acoustic and optic modes)
can be found quite fast numerically (a few minutes for PC even using standard Mathematica software).
Thus we will present results in the form of computed Mathematica plots.

\begin{figure}
\centering
\includegraphics[width=80mm]{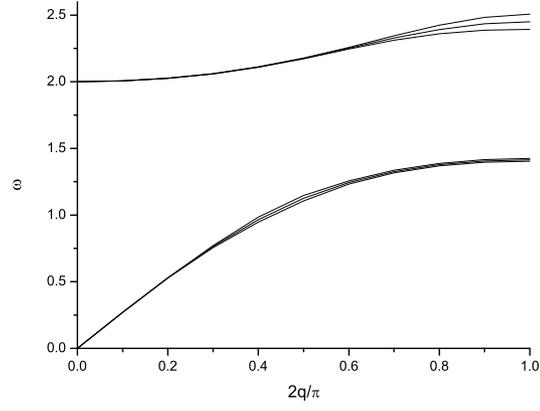}
\caption{Spectrum and broadening for the force constants $\kappa_0=2,\kappa_1=2$.}
\label{f4}
\end{figure}

\begin{figure}
\centering
\includegraphics[width=80mm]{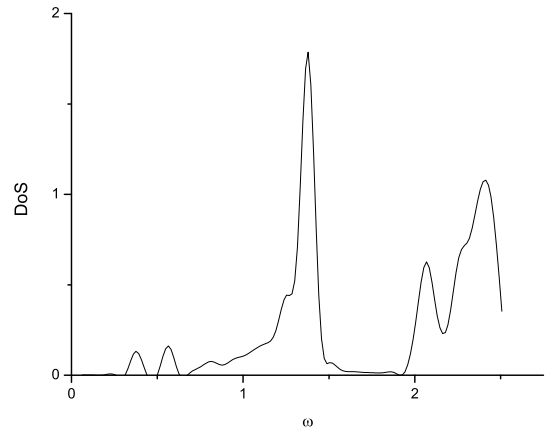}
\caption{Density of states, corresponding to Fig. \ref{f4}.}
\label{f5}
\end{figure}

\begin{figure}
\centering
\includegraphics[width=90mm]{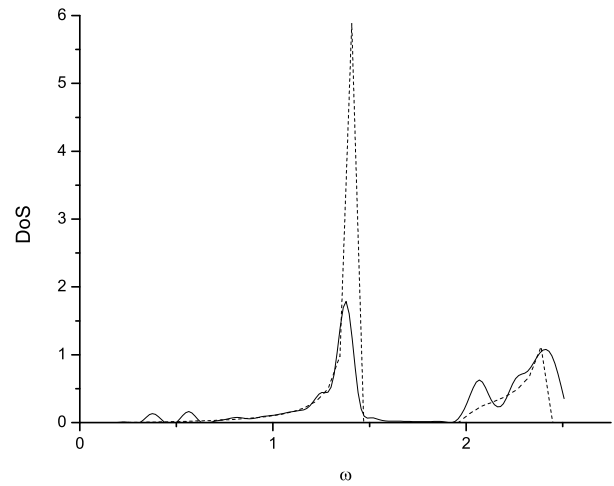}
\caption{Harmonic density of states (dashed lines). Solid line shows the effect of anharmonic smearing on the singularities.}
\label{fnew5}
\end{figure}

\begin{figure}
\centering
\includegraphics[width=90mm]{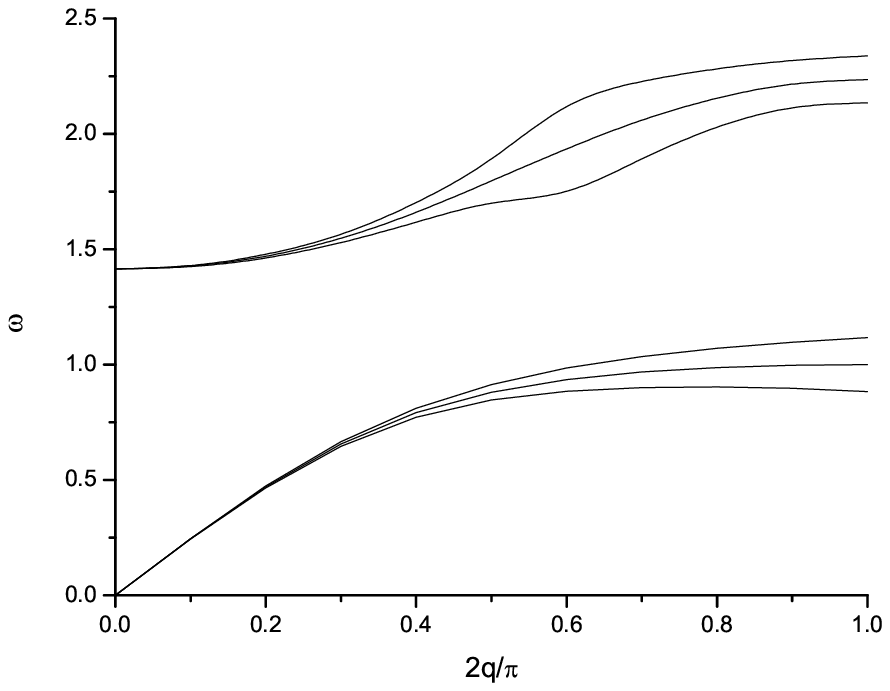}
\caption{Spectrum and broadening for force constants $\kappa_0=2,\kappa_1=1$.}
\label{f6}
\end{figure}

\begin{figure}
\centering
\includegraphics[width=90mm]{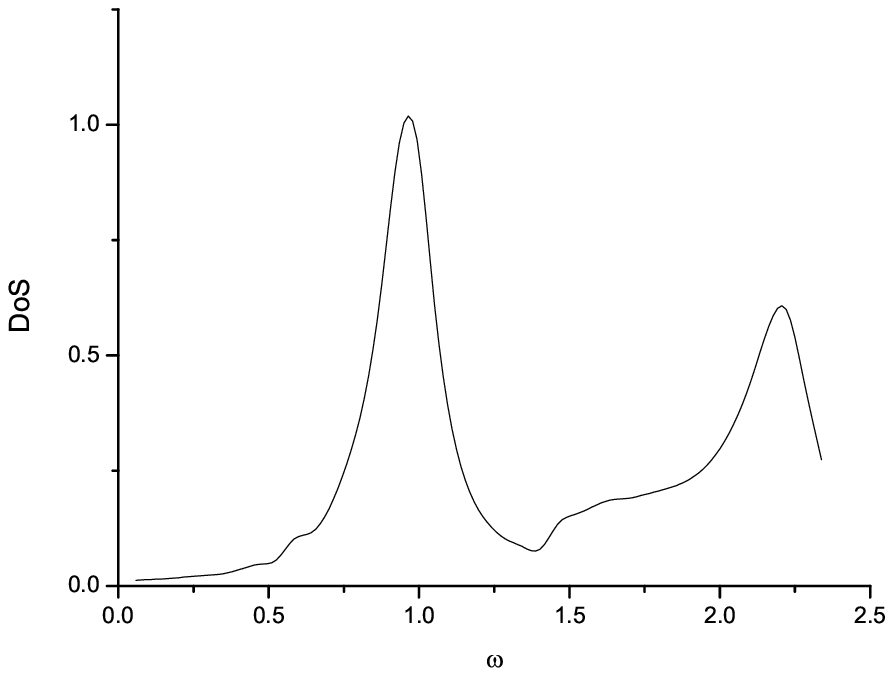}
\caption{Density of states, corresponding to Fig. \ref{f6}.}
\label{f7}
\end{figure}

\begin{figure}
\centering
\includegraphics[width=90mm]{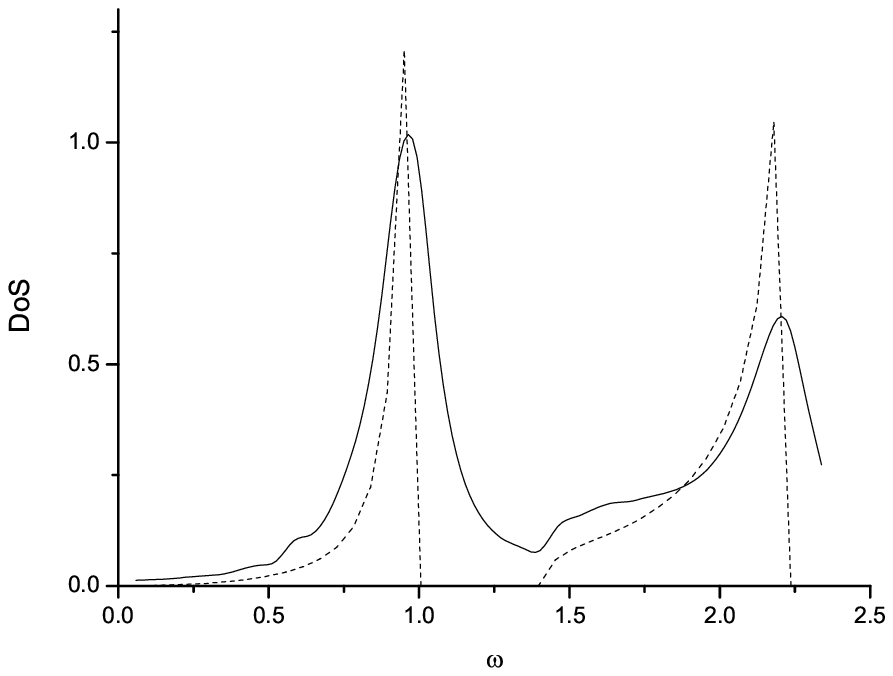}
\caption{Notations are the same as in Fig. \ref{fnew5}.}
\label{fnew7}
\end{figure}

Figure \ref{f4} presents 
the results for the following parameter values $\kappa_0=\kappa_1=2$, $\lambda=0.8$. 
Each branch is presented by three lines, which correspond to $w(q)$ and $w(q)\pm \gamma(q)$.
Figure \ref{f5} shows the calculated (arbitrary normalized) density of states for the same parameter values. 
We should add that  in the low-energy (frequency) region the vibrational density of states has to follow the 
well-known Debye law $\propto \omega ^2$. The wiggles shown in the Fig. \ref{f5} on top of  such a 
parabolic function are due to numerical artifacts. They can be easily
eliminated by merely including more points in the integration code. Since
we are not interested in precise absolute numbers but only in the more or less correct shape of the curves, we restrict ourselves
to only rather crude numerical calculations. 
In the harmonic approximation, the peaks in the vibrational density of states (for the parameters corresponding to Fig. \ref{f5} at $\omega _1 
\simeq 1.5$, and $\omega _2 \simeq 2.5$) transform into  Van Hove singularities (see the Fig. \ref{fnew5}).
Figure \ref{f6} presents the spectra for the parameter values $\kappa_0=2$, $\kappa_1=1$,
and the same triple vertex $\lambda=0.8$. It is easy to see, that the broadening of
the spectrum in this situation where the guest vibrations have softened with respect to the previous case is significantly larger.
Figure \ref{f7} shows the calculated density of states for the same parameter values.
Figure \ref{fnew7} shows (similarly to  Fig. \ref{fnew5}) the harmonic counterpart of the density of states shown in 
Fig. \ref{f7}. 
We should add that the value of the triple vertex used in our calculation was chosen such as to produce a broadening similar to that observed
experimental in the case of soft guest vibrations, 
i.e.\ such that for large wave-vector values the relative width of the phonon line $\gamma/w_1\sim 1/8$.

With all characteristics of the vibrational spectrum in hands, the thermal conductivity of the system $\eta$ can be calculated
by the following expression derived from the Boltzmann kinetic equation \cite{ZI60}, \cite{KI68}
\begin{equation}
\eta= \frac C{6\pi^2} \sum_i \int dq\, q^2\left( \frac{d w_i(q)}{dq}\right) ^2\frac 1{\gamma_i(q)}
\ ,
\label{10}
\end{equation}
where $C$ is the specific heat of the solid and the sum is taken over the two branches of the spectrum.
Unfortunately, bluntly applied, this  in the literature widely used formula for the thermoconductivity $\eta$,
is wrong.
Indeed, in the derivation of this expression it has been assumed that the three-phonon mechanism
of thermal conductivity is operational for all phonon branches, because the energy fluxes carried by the various groups of phonons are additive,
and the same is true for their contributions to  $\eta $. If that mechanism does not work for one group of phonons, it would not
provide finite thermoconductivity. However, the expression (\ref{10}) yields to a finite $\eta $ even if there
were only normal (conserving quasi-momenta) processes, in contradiction to the notorious Peierls theorem \cite{PI29},
claiming that normal processes yield to $\eta = \infty $.
Luckily in our case (isotropic three dimensional guest-host system with two modes with dispersion laws (\ref{4}), this issue
is not relevant. The predominant contributions into the integral (\ref{10}) do not come
from very small wave vectors (excluded by our cut-off procedure), and for wave vectors larger than this cut-off,
just the Umklapp processes limit the thermal flux of higher energy phonons.
The physical picture of heat transfer looks as follows. Normal phonon - phonon collisions, in which the total quasi-momentum is conserved
establish only local internal equilibrium in the phonons, and the Umklapp processes adapt this local equilibrium
to the temperature gradient, and establish a finite value for the thermal conductivity $\eta $.
With said above in mind we are calculating numerically $\eta $ accordingly with Exp. (\ref{10}).

We find that the thermal conductivity in the case, when guest vibrations are soft, turns out significantly smaller, 
than in the pure host system $\eta (\kappa_1=\kappa_0/2)\approx 0.1 \eta (\kappa_1=\kappa_0) $.
This result is independent of the absolute value of the force constants and equally of the anharmonic coupling constant.
Thus, based on our calculations we anticipate that even a relatively modest softening of the guest atom vibrations should
yield a noticeable reduction of thermoconductivity.
We do believe that understanding of the underlying mechanisms is essential to predict the behavior.

\section{Conclusions: New opportunities for studying ?}
\label{con}

Summarizing, in this paper we have investigated certain peculiar features of vibrational dynamics and thermal transport
in guest-host systems. Since these systems are rather complicated and depend on many not easily accessible microscopic 
parameters, any theoretical study has to rely first on very crude simplifications. 
The model investigated theoretically in our paper includes only the minimal ingredients necessary to capture
generic features exhibited by the compounds studied by experiment. We consider only two kinds of particles: strongly coupled
host atoms, which form the structural cavities, and guest atoms confined in these cavities.
In spite of the confinement, for relatively small amplitude vibrations of the guest atoms, their interaction with the host particles
is weaker than between the host atoms.

One of the main difficulties in comparing the results of our effective and simplified model with specific
experimental measurements or elaborated microscopic numerical simulations is the 
availability of an accurate connection between experimental control parameters and 
theoretical model coefficients. The actual values of the parameters are 
determined by the microscopic (atomic level) interactions, and are in fact unknown.
Without prior knowledge of the actual values of microscopic parameters we follow the logic of simplification.
In our model the whole physics is lumped into the effective spring constants $\kappa _0$ and $\kappa _1$ 
and the three phonon interaction parameter $\lambda $.
We calculate the vibrational mode (phonon) dispersion laws, and self-consistently the line broadenings for the empty
and filled (by the guest atoms) cavities. 
We show that the soft phonon sheet (which appears for guest atoms weakly bound in the cavities)
yields large mode broadening via three wave anharmonic mode coupling. 
In turn the Umklapp part of this broadening reduces drastically the phonon thermoconductivity coefficient.
We conclude that for a system with a soft mode vibrational spectrum the broadening is always considerably larger (thermoconductivity is always considerably smaller) than for a system without this soft mode (provided all other characteristic are the same). 

Many of the points made above can be found in the literature, and approaches similar to our paper
have been made by other authors (see e.g. the review paper \cite{KS} and
references therein). However, some important differences to our work should be noted. We focus our study (and investigate in
some detail) on the interplay between softness of the guest mode vibration and the host atom elasticity.
Existing theoretical treatment of thermal conductivity is based on molecular dynamics simulations
\cite{DS01} which should utilize model dependent microscopic interaction potentials, and the Green-Kubo relation for the extraction of the thermal
conductivity from the heat current correlation functions.
Our calculations are much simpler and rely on a model (with only two relevant phenomenological parameters) that is based on general and 
qualitative assumptions. 
We feel that it accommodates a picture of the guest-host system vibrations and thermal transport without going into
finer details. 
Of course we are left with many questions unanswered which must be pursued in further work.
More elaborated models including many particles in the elementary cell, a realistic (corresponding to the crystalline symmetry)
Brillouin zone etc., can be treated within the same approach and with the same conceptual ingredients.
For partially filled cavities in the gust-host systems, special care should be taken about defects
(both, point-like and extended imperfections, like pores and grain-boundaries). Luckily some general conclusions can be obtained without
calculations. First note that while the specific heat and group velocity entering the expression for the thermoconductivity (\ref{10})
are not sensitive to crystal imperfections, the mean free path is. For point defects
which are small compared to the wavelengths of interest,
the phonon scattering cross section $\propto A \omega ^4$, (where the coefficient $A$ is proportional to $V_0(\Delta M/M)^2$
with $V_0$ effective volume of the defect, and $\Delta M$ is the mass difference between the defect, e.g., guest atom,
and the host atom mass $M$). Therefore point 
defects effectively scatter high frequency phonons (however, it should be noted that this expression strictly applies to
a random collection of dilute point defects, whereas the guest molecules are neither dilute nor randomly arranged). 
Because mainly waves of a high group velocity make a substantial contribution to thermal conductivity, in our case
mainly low frequencies play a role. It means that point defects are not very essential for our analysis. Large
size imperfections scatter independently of frequency with a scattering cross section of the order of their
geometrical size. These separation of frequency scales would alow in principle to distinguish
this contribution to the reduction of thermoconductivity from guest-host soft mode mechanism of
the thermoconductivity reduction, investigated in our paper. 
In such a generalized model the precise mode spectrum turns out to be much more complicated than the one found in our paper. 
However, the additional subtleties do not modify the general picture beyond the thermoconductivity reduction. 
Fact is that the parameter that tunes the anharmonic coupling in the high temperature limit $T > T_D$ only scales the final results.
Therefore we anticipate that in the optimum conditions (see section \ref{bas} above), the relative
reduction of the thermal conductivity will be robust with respect to the made approximations and almost $\lambda $
(see the Eqs. (\ref{9})) independent.
In this sense our minimal model is just at the border between those that are too primitive to fit even qualitatively
the data, and those that fit the data too well by using too many parameters.
Our model may be an appropriate tool to generate predictions that can be in principle tested experimentally.


\begin{thebibliography}{99}
\bibitem{KS} 
M.M.~Koza and H.~Schober, {\sl  Vibrational Dynamics and Guest Host
                        Coupling in Clathrate Hydrates} in
                         {\sl Neutron Applications in Earth, Energy and Environmental  Sciences},
                         pp. 351--389, L.~Liang, R.~Rinaldi and H.~Schober (Eds.)  Springer, New York (2008)
\bibitem{SM97} B.C.Sales, D.Mandrus, B.C.Chakoumakos, V.Keppens, J.R.Thompson, Phys. Rev. B, {\bf 56}, 15081 (1997).
\bibitem{DS01} J.Dong, O.F.Sankey, C.W.Myles, Phys. Rev. Lett., {\bf 86}, 2361 (2001).
\bibitem{SI03} H.Schober, H.Itoh, A.Klapproth, V.Chihaia, W.F.Kuhs, Eur. Phys. J. E, {\bf 12}, 41 (2003).
\bibitem{NK08} T.Nakayama, E.Kanashita, Europhys. Lett., {\bf 84}, 66001 (2008).
\bibitem{HB08} D.He, S.Buyukdagli, B.Hu, Phys. Rev. E, {\bf 78}, 061103 (2008).
\bibitem{SA08} K.Suekuni, M.A.Avilas, K.Umeo, H.Fukuoka, S.Yamanaka, T.Nakagawa, T.Takabatake, Phys. Rev. B, {\bf 77}, 235119
(2008).
\bibitem{HJ03} R.P.Hermann, R.Jin, W.Schweika, F.Grandjean, D.Mandrus, B.C.Sales, G.J.Long, Phys. Rev. Lett., {\bf 90}, 135505 (2003).
\bibitem{WH07} H.-C.Wille, R.P.Hermann, I.Sergueev, O.Leupold, P.van der Linden, B.C.Sales, F.Gandjean, G.J.Long,
R.Ruffer, Yu.V.Shvyd'ko, Phys. Rev. B, {\bf 76}, 140301(R) (2007).
\bibitem{SH07} W.Schweika, R.P.Hermann, M.Prager, J.Person, V.Keppens, Phys. Rev. Lett., {\bf 99}, 125501 (2007).
\bibitem{CA08} M.Christensen, A.B.Abrahamsen, N.B.Christensen, F.Juranyi, N.H.Andersen, K.Lefmann, J.Andreasson, C.R.H.Bahl,
Nature Materials, {\bf 7}, 811 (2008).
\bibitem{KJ08} M.M.Koza, M.R.Johnson, R.Viennois, H.Mutka, L.Girard, D.Ravot, Nature Materials, {\bf 7}, 805 (2008).
\bibitem{ZI60} J.M.Ziman, Electrons and Phonos, Claredon Press, Oxford (1960).
\bibitem{KI68} C.Kittel, Introduction to Solid State Physics, Wiley, New York (1968).
\bibitem{HG05} R.P.Hermann, F.Gandjean, G.J.Long, Am. J. Phys., {\bf 73}, 110 (2005).
\bibitem{PI29} R.E.Peierls, Ann. Phys., {\bf 3}, 1055 (1929) (see also the monograph, R.Peierls, Quantum theory of solids, Oxford University
Press, London (1955)).
\bibitem{PO43} I.Pomeranchuk, J.Phys. USSR, {\bf 7}, 197 (1943).
\bibitem{HE54} C.Herring, Phys. Rev., {\bf 95}, 954 (1954).
\bibitem{KL58} P.G.Klemens, Solid State Physics, {\bf 7}, 1 (1958).
\bibitem{OR66} R.Orbach, Phys. Rev. Lett., {\bf 16}, 15 (1966).
\bibitem{KL66} P.G.Klemens, Phys. Rev., {\bf 148}, 845 (1966).
\bibitem{HK93} Y.-J.Han, P.G.Klemens, Phys. Rev. B, {\bf 48}, 6033 (1993).
\bibitem{LH81} M.Lax, P.Hu, V.Narayanamurti, Phys. Rev. B, {\bf 23}, 3095 (1981).
\end{thebibliography}
\end{document}